\documentclass[12pt,a4paper,twocolumn]{article}

\usepackage{subfigure} 
\usepackage{psfig}
\usepackage{chicago}   

\usepackage{graphicx}
\usepackage{epsfig}
\usepackage[latin1]{inputenc}

\makeatletter
\renewcommand{\section}{\@startsection%
{section}%
{1}%
{0mm}%
{- \baselineskip}%
{0.15\baselineskip}%
{\normalfont\normalsize}}%

\renewcommand{\subsection}{\@startsection
{subsection}%
{2}%
{0mm}%
{-\baselineskip}%
{0.15\baselineskip}%
{\normalfont\normalsize}}%
\makeatother

\begin{document}

\title{Transformation of barchans into parabolic dunes under the influence of
vegetation}

\author{\large {O. Dur\'an, V. Schatz, H.
J. Herrmann}\\{\em Institute for Computer Physics, Universit\"at Stuttgart, Germany;}
\\{\large H. Tsoar}\\{\em Ben Gurion University of the Negev, Israel}}
\maketitle
\abstract{
Barchan dunes were found to transform into parabolic dunes and vice
  versa when the amount of vegetation on and around them changes.  This work
  presents the first numerical simulation of this effect.  We propose a
  continuum model for the density of vegetation.  An established sand transport
  model is used for simulating the evolution of the dunes.
}



\section{INTRODUCTION}

Sand dunes are deposits formed by aeolian sand.  They occur frequently in
deserts and on coasts.  The shape of dunes depends on a number of factors, such
as the sand supply, the wind speed, and its directional
variation over the year.  Low sand availability in combination with
unidirectional wind leads to crescent-shaped barchan dunes.

One particular factor which can have a significant influence on the dune shape
is the presence of vegetation on or around the dunes.  A recent investigation
of aerial photographs covering a time span of 50 years \cite{TsoarBlumberg02}
found that barchans can invert their shape to form parabolic dunes and vice
versa when the amount of vegetation changes.  Parabolic dunes are U-shaped
dunes the arms of which point toward the direction of the prevailing wind.  The amount of
vegetation varied over that period because of human activities (such as grazing or 
stabilization) or because of wind power.

The formation of parabolic dunes has been modeled numerically with a lattice
model \cite{Nishimori02}.  Though the authors find intermediate formation
of small parabolic dunes which then form barchans, this does not constitute the
transition between full-sized barchan and parabolic dune found in
\cite{McKee71,TsoarBlumberg02}.  This effect has not been investigated theoretically
before.  We propose a model for vegetation growth taking into account sand
erosion and deposition and use an established saltation model for simulating
the sand transport which determines the evolution of the dunes.

\section{MODELS}

\begin{figure*}[!ht]
\includegraphics[width=1.0 \textwidth]{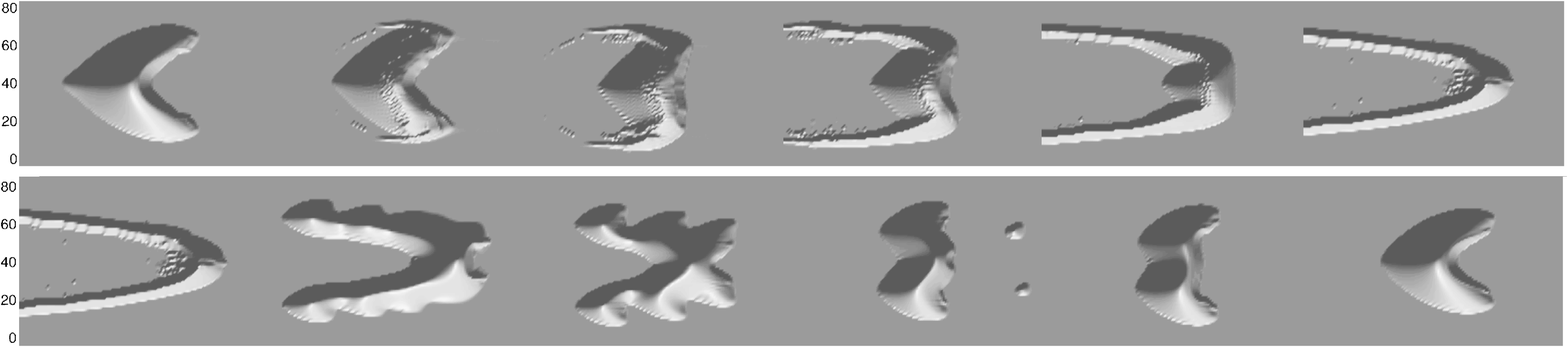}
\caption{Evolution of an initial barchan dune to a fixed parabolic shape
and
vice versa. For the first transformation we allowed that the vegetation grows
with a characteristic growth time ($\tau_s$) of 7 days
(the evolution of vegetation appears in figure \ref{veget-fig}).
Whereas for the second, the vegetation is removed and the dune is left to evolve normally.}
\label{trans-fig}
\end{figure*}
\begin{figure*}
\includegraphics[width=1.0 \textwidth]{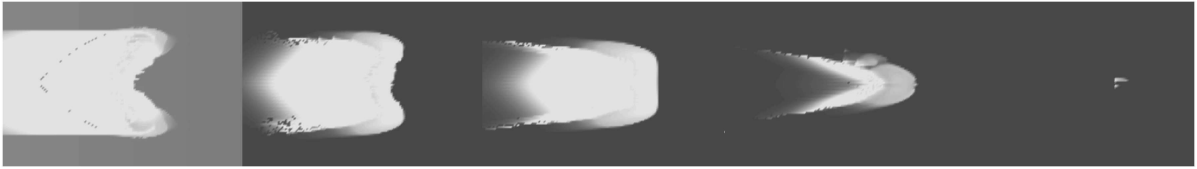}
\caption{Evolution of the vegetation density in a barchan-parabolic
transformation for a
vegetation characteristic growth time of 7 days (see upper part
of Figure \ref{trans-fig}). The gray value represents the vegetation density:
black means
complete cover, white, no vegetation.}
\label{veget-fig}
\end{figure*}

\subsection{Vegetation Growth:}
We characterize the vegetation by its local height ($h_v$) and,
for the vegetation growth rate, we propose a continuous model inspired by
\cite{Nishimori02}.

We suppose, first, that vegetation can grow until it reaches a
maximum height ($H_v$) and, second, that the growth process has a characteristic
time ($\tau_s$),
determined by
 climatic conditions that enhance or inhibit it.

Moreover, the
vegetation growth rate should
 be a function of the time rate of sand surface change ($\partial h/\partial
t$). After any temporal change
of the sand surface ($h$) the vegetation needs time to adapt to the new
conditions. We introduce this effect only as a delay in the vegetation growth.
Thus:

\begin{equation}
\label{vegevol}
\frac{dh_v}{dt} = \frac{H_v - h_v}{\tau_s} - \left | \frac{\partial h}{\partial
t} \right |
\end{equation}

However, equation \ref{vegevol} is not enough for describing the evolution of
vegetation. Under a continuous sand erosion the vegetation dies because its
roots are exposed \cite{TsoarBlumberg02}
. In this case what is important is
the total erosion of the sand bed and not
its time rate. Thus, we choose the following criterion: if the local sand level
is reduced by more than 10\% of the vegetation
height above it, then the plant dies. This parameter just takes into account a limit for the exposition of
the plant's root and the results does not depend on its exact value.

Another important assumption is related to the places at which vegetation can
grow. We considered that in those places that
have been covered by sand the vegetation cannot grow before a time interval
($t_v$) necessary for the soil recovery. We choose this time interval as 8 months, approximately the
turn over time of the Barchan dune we use in the simulation.

\subsection{Shear Stress Partitioning:}

The shear stress partitioning is the main dynamical effect of the vegetation
on the flow field and, hence, on the sand transport. The
vegetation acts as
roughness that
absorb part of the momentum transfered to the soil by the wind, thus,
the total surface shear stress is divided into two components, one acting on vegetation and the
other on the sand
grains. The
fraction of the total
shear stress acting on the sand grains can be described by the expression
\cite{Buckley87}:

\begin{equation}
\label{tauveg}
\tau_s = \left( 1- \frac{\rho_v}{\rho_c} \right)^2 \, \tau
\end{equation}

\noindent
where $\tau$ is the total surface shear stress, $\tau_s$ is the shear stress
acting on the
non-vegetated ground,
$\rho_v$ is the vegetation density, defined as $(h_v/H_v)^2$, and $\rho_c$ is
a critical vegetation density that
depends mainly on the geometric properties of the vegetation \cite{PyeTsoar90}.

Equation \ref{tauveg} represents a reduction of the shear stress acting on the
sand  grains that also implies a reduction of the sand flux. Both equations
(\ref{vegevol} and \ref{tauveg}) contain the interaction between the
vegetation  and the sand surface. In those places where the sand
erosion or deposition is small enough the vegetation grows (\ref{vegevol}).
Then, the shear stress, and also the sand flux, decreases (\ref{tauveg}) and
starts the sand deposition, which in turn slows down the vegetation growth.

\subsection{Sand Transport:}
For the dune evolution we use an established sand transport model \cite{SauerKroy01,Schwaemmle03} that
consists of three coupled equations for the wind shear stress, the sand flux
and the avalanches, and the resulting change in the sand surface using mass
conservation.

\section{RESULTS}

We performed simulations placing a $4.2\,m$ high barchan dune on a rock
bed and then
allowing the vegetation to grow. A zero influx and
a $0.5\,m/s$ upwind shear velocity are set.
We also fixed $\rho_c= 0.5$, a typical value for
spreading herbaceous dune plants \cite{PyeTsoar90} and $H_v=1.0\,m$.
Finally, the dune model parameters have been specified in \cite{Sauermannphd}.

We studied the influence of the
vegetation characteristic growth time ($\tau_s$) which contains the information
of the growth rate and, hence, controls the strength of the interaction between
the dune
and the vegetation.

The upper part of Figure \ref{trans-fig} shows snapshots of the evolution
of a barchan dune under the influence of vegetation with a characteristic
growth time of 7 days.  The evolution of the vegetation density is shown in
Figure \ref{veget-fig}. Initially, the vegetation invades those places where
the sand erosion or deposition is small, the horns, the crest and the surroundings
of the dune except upwind. There, the soil was covered by sand, and, as a consequence
of our model, it needs a time ($\tau_s$) to recover.
As the vegetation grows, it traps the sand, which then cannot reach the lee side. There,
the
vegetation cover increases.
On the other hand, the vegetation on the windward side is
eliminated because its roots are exposed as the dune migrates.
However, at the horns the vegetation grows fast enough to survive the small
sand deposition and, there, the sand accumulates. Hence, whereas the central
part of the dune moves forward a sand trail is left behind at the horns. This
process leads to the stretching of the windward side and a
formation of a parabolic dune.

This picture agrees well with a recent conceptual model to explain such transformation
based
on field observations \cite{TsoarBlumberg02}.

The bottom of figure \ref{trans-fig} shows the inverse process. After
eliminating all vegetation and setting a constant influx of $0.005 \,kg/ms$,
the parabolic dune is fragmented into small barchanoid forms that nucleate
into a final barchan dune. Although the transformation from parabolic
into barchan or transversal dunes have been observed \cite{Anton86}, this fragmentation does
not occur in natural conditions because the vegetation is eliminated gradually from the
parabolic dune, leaving first the central part but still remaining at the
lateral trails.

\begin{figure}[!t]
\begin{center}
\includegraphics[width=0.5 \textwidth]{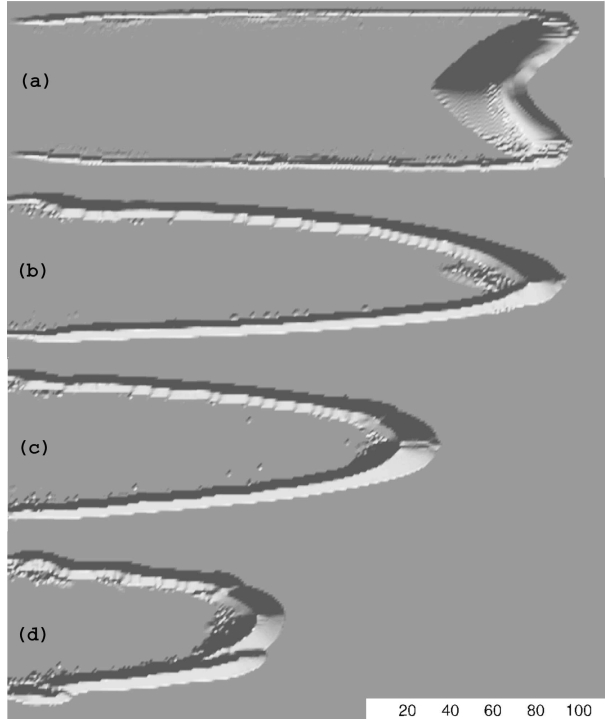}
\caption{Final or stationary states starting with one barchan dune under the
influence of vegetation with
different growth rates.The vegetation characteristic growth times are:
(a) 12, (b) 8, (c) 7 and (d) 4.6 days. The parabolic dunes are (b) $260\,m$, (c) $200\,m$ and
(d) $130\,m$ long.
Note that in (a) the vegetation cover is
not dense enough
to transform a barchan dune into a parabolic one.}
\label{parab-fig}
\end{center}
\end{figure}

The barchan to parabolic transition and also the parabolic shape, strongly
depend on the vegetation
growth rate.
Figure \ref{parab-fig} (b), (c) and (d) show the final parabolic dune
for three values of $\tau_s$ (see figure caption). As we expected a longer
parabolic dune emerges for a smaller vegetation growth rate, i.e. a higher $\tau_s$.
On the other hand, if $\tau_s$ is too high ,the vegetation cover is not enough to complete
the inversion process (Figure \ref{parab-fig} (a)).
In this case the barchan keeps its shape but leaves lateral sand trails covered by plants. However, due to 
the constant loss of sand in the arms, the Barchan size is reduced until finally being stabilized 
by vegetation.

Notice that the parabolic dunes are slightly asymmetric (Figure \ref{parab-fig}) despite
the initial condition being symmetric. This surprising result is the consequence of the
interaction of the
vegetation with the sand bed. Once vegetation grows it protects the soil from erosion.
This enhances the growth process, which in turn, increments the soil protection and so on.
This mechanism amplifies small asymmetries in the vegetation cover. In our simulation, the
initial small asymmetries are due to numerical inaccuracies, in nature, they are a consequence
of random factors influencing vegetation growth. When the
 vegetation cover is small or zero these asymmetries disappear.

\section{CONCLUSIONS}

We performed a numerical simulation of the influence of varying amounts of
vegetation on dune shapes.  We proposed a continuum model describing vegetation
growth (Equation~\ref{vegevol}).  Taking into account the partitioning of the
shear stress between the plants and the ground, we used a saltation model to
simulate the evolution of the dune shape.

We have reproduced the observed effect of the transition between barchans and
parabolic dunes.  When vegetation is allowed to grow on a barchan according to
(Equation \ref{vegevol}), it inverts its curvature to become a parabolic dune.  When the
vegetation is removed, it transforms back into a barchan.  In the latter
process, we apply a small sand influx to compensate for the loss of sand which
is no longer trapped by the vegetation.

After a parabolic dune is formed, it is completely covered by vegetation and
rendered inactive.  We have found that the final shape of the parabolic dune
evolving from a barchan depends strongly on the growth rate of the vegetation.
Slow-growing plants only slow down the arms of the barchan and do not transform
it completely into a parabolic dune.  The faster the plants grow, the faster
the transformation is completed and the shorter are the arms of the resulting
parabolic dune.

\bibliography{dune}

\end{document}